%
\let\useblackboard=\iftrue
%
\let\useblackboard=\iffalse
%
%
\input harvmac.tex

%
\input epsf.tex
\ifx\epsfbox\UnDeFiNeD\message{(NO epsf.tex, FIGURES WILL BE
IGNORED)}
\def\figin#1{\vskip2in}
\else\message{(FIGURES WILL BE INCLUDED)}\def\figin#1{#1}\fi
\def\ifig#1#2#3{\xdef#1{fig.~\the\figno}
\midinsert{\centerline{\figin{#3}}%
\smallskip\centerline{\vbox{\baselineskip12pt
\advance\hsize by -1truein\noindent{\bf Fig.~\the\figno:} #2}}
\bigskip}\endinsert\global\advance\figno by1}
\noblackbox
\baselineskip=12pt
\useblackboard
\message{If you do not have msbm (blackboard bold) fonts,}
\message{change the option at the top of the tex file.}

\font\blackboard=msbm10 scaled \magstep1
\font\blackboards=msbm7
\font\blackboardss=msbm5
\textfont\black=\blackboard
\scriptfont\black=\blackboards
\scriptscriptfont\black=\blackboardss

\else

\fi

\def\lsim{\mathrel{\lower3pt\hbox{$\sim$}}
\hskip-11.5pt\raise1.6pt\hbox{$<$}\;}

\def\gsim{\mathrel{\lower3pt\hbox{$\sim$}}
\hskip-11.5pt\raise1.6pt\hbox{$>$}\;}

\def\yboxit#1#2{\vbox{\hrule height #1 \hbox{\vrule width #1
\vbox{#2}\vrule width #1 }\hrule height #1 }}
\def\fillbox#1{\hbox to #1{\vbox to #1{\vfil}\hfil}}
\def\ybox{{\lower 1.3pt \yboxit{0.4pt}{\fillbox{8pt}}\hskip-0.2pt}}

\def\comments#1{}

\def\Im{{\rm Im\hskip0.1em}}

\def\Pslash#1{\rlap{\hskip0.2em/}{#1}}

\def\CN{{\cal N}}

\def\II{\relax{I\kern-.07em I}}

\def\IIB{{\II}B}

\def\IZ{\relax\ifmmode\mathchoice
{\hbox{\cmss Z\kern-.4em Z}}{\hbox{\cmss Z\kern-.4em Z}}
{\lower.9pt\hbox{\cmsss Z\kern-.4em Z}}
{\lower1.2pt\hbox{\cmsss Z\kern-.4em Z}}\else{\cmss Z\kern-.4em
Z}\fi}

\def\Im{{\rm Im\ }}

\def\nfour{$\CN \!\!=\!\! 4$}

\def\none{$\CN \!=\! 1$}
%
%

%
%
\def\pnc{p_{nc}}
%
%
%
%
%
%
\lref\sus{
A.~Matusis, L.~Susskind and N.~Toumbas,
``The IR/UV connection in the non-commutative gauge theories,''
JHEP{\bf 0012} (2000) 002
[hep-th/0002075].
}
\lref\haya{
M.~Hayakawa,
``Perturbative analysis on infrared and ultraviolet aspects of  noncommutative QED on R**4,''
hep-th/9912167.
}
\lref\gubser{
S.~S.~Gubser and S.~L.~Sondhi,
``Phase structure of non-commutative scalar field theories,''
hep-th/0006119.
}
\lref\ruiz{
F.~R.~Ruiz,
``Gauge-fixing independence of IR divergences in non-commutative U(1),  perturbative tachyonic instabilities and supersymmetry,''
Phys.\ Lett.\ B {\bf 502} (2001) 274
[hep-th/0012171].
}

\lref\hooft{
G.~'t Hooft and M.~Veltman,
``Combinatorics Of Gauge Fields,''
Nucl.\ Phys.\ B {\bf 50} (1972) 318;
B.~W.~Lee and J.~Zinn-Justin,
``Spontaneously Broken Gauge Symmetries. I. Preliminaries,''
Phys.\ Rev.\ D {\bf 5} (1972) 3121;
D. Gross, ``Applications of the Renormalization Group to High-Energy Physcs,''
in {\it Methods in field theory}, edited by R. Balian and J. Zinn-Justin (North-Holland, New York, 1976).
}

\lref\strominger{
R.~Gopakumar, S.~Minwalla and A.~Strominger,
``Noncommutative solitons,''
JHEP{\bf 0005} (2000) 020
[hep-th/0003160].
}
\lref\we{K. Landsteiner, E. Lopez and M. H. Tytgat,
``Excitations in hot non-commutative theories,''
JHEP{\bf 0009}, 027 (2000)
[hep-th/0006210].
}
\lref\kapusta{J. Kapusta, ``Finite temperature field theory'', Cambridge
Univ. Press, 1990.}
\lref\bellac{M. Le Bellac, ``Thermal field theory'', Cambridge Univ. Press,
1996.}
\lref\rob{
R.~D.~Pisarski,
``Scattering Amplitudes In Hot Gauge Theories,''
Phys.\ Rev.\ Lett.\ {\bf 63} (1989) 1129;
E.~Braaten and R.~D.~Pisarski,
``Soft Amplitudes In Hot Gauge Theories: A General Analysis,''
Nucl.\ Phys.\ B {\bf 337} (1990) 569.
}
\lref\rebhan{
R.~Kobes, G.~Kunstatter and A.~Rebhan,
``Gauge dependence identities and their application at finite temperature,''
Nucl.\ Phys.\ B {\bf 355} (1991) 1.
}
\lref\sw{
N.~Seiberg and E.~Witten,
``String theory and noncommutative geometry,''
JHEP{\bf 9909} (1999) 032
[hep-th/9908142].
}
\lref\aki{
A.~Hashimoto and N.~Itzhaki,
``Traveling faster than the speed of light in non-commutative geometry,''
hep-th/0012093.
}
\lref\park{
D.~Bak, K.~Lee and J.~Park,
``Noncommutative vortex solitons,''
hep-th/0011099.
}
\lref\csaba{
D.~J.~Chung, E.~W.~Kolb and A.~Riotto,
``Extra dimensions present a new flatness problem,''
hep-ph/0008126;

C.~Csaki, J.~Erlich and C.~Grojean,
``Gravitational Lorentz violations and adjustment of the cosmological  constant in asymmetrically warped spacetimes,''
hep-th/0012143.
}
\lref\langlois{
R.~R.~Caldwell and D.~Langlois,
``Shortcuts in the fifth dimension,''
gr-qc/0103070.
}
\lref\riotto{
L.~Pilo and A.~Riotto,
``The non-commutative brane world,''
hep-ph/0012174.
}
\lref\martin{
C.~P.~Martin and D.~Sanchez-Ruiz,
``The BRS invariance of noncommutative U(N) Yang-Mills theory at the  one-loop level,''
Nucl.\ Phys.\ B {\bf 598} (2001) 348
[hep-th/0012024].
}
\lref\mvrs{
S.~Minwalla, M.~Van Raamsdonk and N.~Seiberg,
``Noncommutative perturbative dynamics,''
JHEP{\bf 0002} (2000) 020
[hep-th/9912072].
}
\lref\gkf{
A.~Gonzalez-Arroyo and C.~P.~Korthals Altes,
``Reduced Model For Large N Continuum Field Theories,''
Phys.\ Lett.\ B {\bf 131} (1983) 396;

T.~Filk,
``Divergencies in a field theory on quantum space,''
Phys.\ Lett.\ B {\bf 376} (1996) 53.
}

\lref\zan{
M.~Pernici, A.~Santambrogio and D.~Zanon,
``The one-loop effective action of noncommutative N = 4 super Yang-Mills  is gauge invariant,''
hep-th/0011140;
A.~Santambrogio and D.~Zanon,
``One-loop four-point function in noncommutative N=4 Yang-Mills theory,''
JHEP{\bf 0101} (2001) 024
[hep-th/0010275].
}

\lref\gkw{
H.~Grosse, T.~Krajewski and R.~Wulkenhaar,
``Renormalization of noncommutative Yang-Mills theories: A simple  example,''
hep-th/0001182;
I.~Chepelev and R.~Roiban,
``Convergence theorem for non-commutative Feynman graphs and  renormalization,''
JHEP{\bf 0103} (2001) 001
[hep-th/0008090].
}

\lref\ggrs{
H.~O.~Girotti, M.~Gomes, V.~O.~Rivelles and A.~J.~da Silva,
``A consistent noncommutative field theory: The Wess-Zumino model,''
Nucl.\ Phys.\ B {\bf 587} (2000) 299
[hep-th/0005272].
A.~A.~Bichl, J.~M.~Grimstrup, H.~Grosse, L.~Popp, M.~Schweda and R.~Wulkenhaar,
``The superfield formalism applied to the noncommutative Wess-Zumino  model,''
JHEP{\bf 0010} (2000) 046
[hep-th/0007050].
}

\lref\akione{
A.~Hashimoto and N.~Itzhaki,
``Non-commutative Yang-Mills and the AdS/CFT correspondence,''
Phys.\ Lett.\ B {\bf 465} (1999) 142
[hep-th/9907166].
}
\lref\maldacena{
J.~M.~Maldacena and J.~G.~Russo,
``Large N limit of non-commutative gauge theories,''
JHEP{\bf 9909} (1999) 025
[hep-th/9908134].
}
\lref\harvey{
J.~A.~Harvey,
``Komaba lectures on noncommutative solitons and D-branes,''
hep-th/0102076.
}

\lref\bigsus{
D.~Bigatti and L.~Susskind,
``Magnetic fields, branes and noncommutative geometry,''
Phys.\ Rev.\ D {\bf 62} (2000) 066004
[hep-th/9908056].
}
\lref\petriello{
F.~J.~Petriello,
``The Higgs mechanism in non-commutative gauge theories,''
hep-th/0101109.
}

\lref\avm{G.~Arcioni and M.~A.~Vazquez-Mozo,
``Thermal effects in perturbative noncommutative gauge theories,''
JHEP{\bf 0001} (2000) 028
[hep-th/9912140];
W.~Fischler, E.~Gorbatov, A.~Kashani-Poor, S.~Paban, P.~Pouliot and J.~Gomis,
``Evidence for winding states in noncommutative quantum field theory,''
JHEP{\bf 0005} (2000) 024
[hep-th/0002067];
W.~Fischler, E.~Gorbatov, A.~Kashani-Poor, R.~McNees, S.~Paban and P.~Pouliot,
``The interplay between Theta and T,''
JHEP{\bf 0006} (2000) 032
[hep-th/0003216];
G.~Arcioni, J.~L.~Barbon, J.~Gomis and M.~A.~Vazquez-Mozo,
``On the stringy nature of winding modes in noncommutative thermal field  theories,''
JHEP{\bf 0006} (2000) 038
[hep-th/0004080].
J.~Gomis, K.~Landsteiner and E.~Lopez,
``Non-relativistic non-commutative field theory and UV / IR mixing,''
Phys.\ Rev.\ D {\bf 62}, 105006 (2000)
[hep-th/0004115].
}

\lref\miguel{
G.~Arcioni and M.~A.~Vazquez-Mozo as in \avm.}

\lref\gpy{
D.~J.~Gross, R.~D.~Pisarski and L.~G.~Yaffe,
``QCD And Instantons At Finite Temperature,''
Rev.\ Mod.\ Phys.\ {\bf 53} (1981) 43.
}
\lref\ncos{
S.~S.~Gubser, S.~Gukov, I.~R.~Klebanov, M.~Rangamani and E.~Witten,
``The Hagedorn transition in non-commutative open string theory,''
hep-th/0009140.
}

\lref\plasmi{
H.~A.~Weldon,
``Dynamical holes in the quark-gluon plasma,''
Phys.\ Rev.\ {\bf D40} (1989) 2410.
}

\lref\wess{
B.~Jurco, P.~Schupp,
``Noncommutative Yang-Mills from equivalence of star products,''
Eur.\ Phys.\ J.\ C {\bf 14} (2000) 367, [hep-th/0001032].
B.~Jurco, P.~Schupp, J.~Wess,
``Noncommutative gauge theory for Poisson manifolds,''
Nucl.\ Phys.\ B {\bf 584} (2000) 784 [hep-th/0005005].
}
\lref\ncall{
A.~Connes, M.~R.~Douglas and A.~Schwarz,
``Noncommutative geometry and matrix theory: Compactification on tori,''
JHEP{\bf 9802} (1998) 003
[hep-th/9711162];
M.~R.~Douglas and C.~Hull,
``D-branes and the noncommutative torus,''
JHEP{\bf 9802} (1998) 008
[hep-th/9711165];
Y.~E.~Cheung and M.~Krogh,
``Noncommutative geometry from 0-branes in a background B-field,''
Nucl.\ Phys.\ B {\bf 528} (1998) 185
[hep-th/9803031];
C.~Chu and P.~Ho,
``Noncommutative open string and D-brane,''
Nucl.\ Phys.\ B {\bf 550} (1999) 151
[hep-th/9812219];
F.~Ardalan, H.~Arfaei and M.~M.~Sheikh-Jabbari,
``Noncommutative geometry from strings and branes,''
JHEP{\bf 9902} (1999) 016
[hep-th/9810072].
}
\lref\pr{
J.~L.~Barbon and E.~Rabinovici,
``On the nature of the Hagedorn transition in NCOS systems,''
hep-th/0104169.
}
\lref\arm{
A.~Armoni,
``Comments on perturbative dynamics of non-commutative Yang-Mills theory,''
Nucl.\ Phys.\ B {\bf 593} (2001) 229
[hep-th/0005208].
}
\lref\lp{
L.~Alvarez-Gaume, J.~L.~Barbon and R.~Zwicky,
``Remarks on time-space noncommutative field theories,''
hep-th/0103069.
}
\lref\kh{
T.~J.~Hollowood, V.~V.~Khoze and G.~Travaglini,
``Exact results in noncommutative N = 2 supersymmetric gauge theories,''
hep-th/0102045.
}

%
\Title{ \vbox{\baselineskip12pt\hbox{hep-th/0104133}
\hbox{CERN-TH-2001-106}\hbox{ULB-TH/01-09}
}}
{\vbox{Instability of Non-Commutative SYM Theories
\centerline{at Finite Temperature} }}
\centerline{Karl Landsteiner, Esperanza Lopez}
\medskip
\centerline{Theory Division CERN}
\centerline{CH-1211 Geneva 23}
\centerline{Switzerland}
\medskip
\centerline{Michel H.G. Tytgat}
\medskip 
\centerline{Service de Physique Th\'eorique, CP225}
\centerline{Universit\'e Libre de Bruxelles}
\centerline{Bld du Triomphe, 1050 Brussels, Belgium}
\medskip
\centerline{\tt Karl.Landsteiner@cern.ch}
\centerline{\tt Esperanza.Lopez@cern.ch}
\centerline{\tt mtytgat@ulb.ac.be}

\vskip15mm

\centerline{\bf Abstract}

\baselineskip=14pt
We extend our previous work on the quasi-particle excitations in \nfour\ 
non-commutative $U(1)$ Yang-Mills theory at finite temperature. 
We show that above some critical 
temperature there is a tachyon in the spectrum of excitations. 
It is a collective transverse photon mode 
polarized in the non-commutative plane.
Thus the theory seems to undergo a phase transition at high temperature.
Furthermore we find that the group velocity of quasi-particles generically
exceeds the speed of
light at low momentum.

\vfill
\Date{\vbox{\hbox{\sl April 2001}}}

\newsec{Introduction}

Quantum field theories on non-commutative spaces exhibit the intriguing 
phenomenon of UV/IR mixing \mvrs. On a technical level this arises
as follows. Like in large $N$ gauge theories, one  distinguishes between 
planar and non-planar graphs. Planar graphs have the usual UV behaviour of 
ordinary quantum field theories because the phases stemming from the Moyal 
product do not depend on the internal momenta. In non-planar graphs 
the phases do depend on the internal momenta \gkf, leading in general
to the regularization of potentially divergent integrals. This 
regularization depends however on the inflowing momentum. 
The UV divergence reappears in the disguise of an IR divergence as the
inflowing momentum goes to zero. 
This implies that high momentum modes do not decouple 
from the physics at large distances and furthermore makes 
renormalization of non-commutative field theories a difficult issue.

Depending on the details of the theory, UV/IR mixing may also lead to the
appearance of tachyonic modes. This has been shown already in \mvrs, in the 
case of a $\lambda \phi^3$ theory in six dimensions. Because the potential is 
unbounded from below, this theory is expected to be unstable.
However, in the ordinary case the instability is non-perturbative in a weak 
coupling expansion, while in the non-commutative case an instability 
appears already at one-loop. Another example of a theory that can
be unstable at one-loop due to UV/IR mixing is the non-commutative version 
of a $\;U(1)$ gauge theory. The photon polarization tensor has been first 
calculated in \haya\sus, where it was shown that it acquires
generically a pole-like IR divergent piece. The nature of the IR pole 
depends on the content of adjoint matter. It is remarkable that,
if there are more bosonic than fermionic fields in the adjoint
representation, the IR pole leads to the appearance of tachyonic modes
at long wavelengths. In particular, tachyonic modes appear at one-loop
level for pure $U(1)$ gauge theory. In \ruiz\ it was shown
that the tachyonic character of the IR pole is gauge independent. 
As of today, a satisfactory understanding of the effects of UV/IR mixing 
in general and of the existence of instabilities in particular is still
missing. 

Pole-like infrared divergences are absent in supersymmetric theories \sus,
and thus the implications of UV/IR mixing are somewhat less severe. 
Indeed, it has been argued that UV/IR mixing effects in logarithmic 
divergences do not threaten renormalizability \gkw\ggrs. 
In four space-time dimensions, the maximally supersymmetric theory is 
\nfour\ Yang-Mills theory. The commutative version is known to be UV 
finite and therefore one expects the non-commutative theory to be free of 
both pole-like and logarithmic IR divergences. This has been shown at 
one-loop level in \zan. Non-commutative \nfour\ Yang-Mills theory is 
also interesting from the point of view of string theory. It arises as 
the effective low energy theory describing the physics of D3-branes in 
\IIB\ string theory in the background of a magnetic B field \ncall.  
Many features of non-commutative gauge theories, like for instance 
non-commutative solitons, have a natural D-brane interpretation. 
(See {\it e.g.} \harvey\ for a review.)

Since perturbation theory seems 
well-defined in \nfour\ non-commutative
Yang-Mills theory it is  natural to ask what happens when  supersymmetry 
is broken. One possible way to investigate this problem is by 
introducing temperature. Non-commutative theories
at finite temperature have already been considered in a number of earlier
works \avm\we. In particular, in \we\ we argued that introducing finite 
temperature allows to study the issue of UV/IR mixing in
a controllable setting. Of course, 
conventional Yang-Mills field theories at finite temperature
have their own share of IR troubles. (See {\it e.g.} \gpy\ or \kapusta.)  
However, if one considers non-commutative theories with Moyal 
bracket interactions, these specific problems are absent, simply because 
interactions switch-off for vanishing momentum. Therefore, it seems that 
\nfour\ non-commutative $\;U(1)$ at finite temperature is an 
interesting starting point to study the issue of UV/IR mixing. 

In this paper we will extend our previous work on scalar dispersion relations
in supersymmetric, non-commutative field theories. Specifically we will
investigate the dispersion relation of vector fields and fermions in
\nfour\ non-commutative $\;U(1)$ gauge theory at the one-loop level. 
The work is organized as follows.
In section 2 we review our previous findings on scalar dispersion relations.
We discuss the appearance of superluminous wave propagation at the lower
end of the spectrum. We discuss the issue of causality and the relation
to the string picture where the non-commutative theory appears in the
Seiberg-Witten decoupling limit of a D3-brane in a background magnetic field. 

Section 3 is the main body of the paper. We compute the polarization 
tensor of the vector bosons in the so-called hard thermal loop approximation 
(HTL) \rob.
Compared to conventional gauge theories at finite temperature, 
\nfour\ non-commutative $\;U(1)$ presents a number of remarkable features.
We show that 
a longitudinal collective excitation (the non-commutative relative of the 
plasmon mode of standard gauge theories) appears  above a certain critical
temperature $T_c$. The existence of a critical temperature is a new feature 
which can be understood intuitively by picturing non-commutative particles 
in the adjoint as rigid dipoles \bigsus. The presence of a thermal bath 
also affects 
the transverse modes. In our setup we have to distinguish between transverse 
states polarized respectively in the commutative and non-commutative 
directions. Our most striking result is that, above $T_c$, there is a new 
collective transverse mode polarized in the non-commutative plane of 
tachyonic nature. We observe that, as $T\!\rightarrow \!\infty$ and in the 
static limit, the
dispersion relations we find reproduce the pole-like IR divergence
of the reduced three-dimensional non-supersymmetric theory which is the 
high temperature limit of the \nfour\ theory.  

In section 4 we complete our work by computing the self-energy of the 
fermions. At finite temperature generically we expect collective fermion
excitations. Again we find that these are present in the model only for 
high temperatures. The critical temperature for the existence or
these plasmino modes turns out to be higher than for the collective 
vector modes. 

In section 5 we discuss our results and speculate on the physical implications
of the tachyon. In two appendices we have collected the expressions for 
the polarization tensor and the fermion self-energy that result from the 
expansion of the Moyal phases, and give a short discussion of gauge 
independence of the two-point functions in the HTL approximation.

Finally let us set the stage for our calculation by introducing some 
conventions. Non-commutativity of the coordinates takes the form
\eqn\ncomm{ [x^\mu, x^\nu ] = i \theta^{\mu\nu}\, .}
Time is taken to be an ordinary commuting coordinate and $\theta^{\mu\nu}$ 
is an $x$-independent matrix.
The algebra of functions is defined by the Moyal product
\eqn\moyal{ f(x) * g(x) = \lim_{y\rightarrow x}e^{i \theta^{\mu\nu}
\partial^x_\mu \partial^y_\nu } f(x) g(y) \,.}
Star product commutators or so-called Moyal brackets are
defined as $\{f,g\}_* (x)$  $= f(x)*g(x) - g(x)*f(x)$. The non-commutative
field strength is $F_{\mu\nu} = \partial_\mu A_\nu -  \partial_\nu A_\mu-
i \{A_\mu,A_\nu\}_*$.
The perturbative expansion is most easily derived by considering a 
ten-dimensional \none\ theory 
\eqn\action{ S = {1\over 2 g^2} \int d^{10}x \left(-{1\over 2} F^{\mu\nu} F_{\mu\nu} +
\bar\psi \Gamma^\mu ( i \partial_\mu \psi + \{A_\mu,\psi\}_* )\right) ,}
and restricting the momenta to four dimension. 
We will use capital letters to refer to four-momentum.
In Euclidean signature we will have $p_0 = i \omega$.
Contraction of a vector with $\theta^{\mu\nu}$ will be denoted by 
$\tilde{p}^\mu = \theta^{\mu\nu} p_\nu$ 
and spatial momenta pointing in the non-commuting directions noted $\pnc$.
Finally, without loss of generality we will 
assume $\theta^{23}=\theta$, with all other components set to zero. 
The calculations are done using the Feynman gauge. 
(See also Appendix B.)

\newsec{Superluminal Propagation} 

We first review the results of \we\ on the dispersion relation of
scalar excitations in non-commutative \nfour $\;U(1)$ gauge theory at 
finite temperature.
The scalar self-energy at one-loop can be written in the form
\eqn\selfscalar{\Sigma = 32 g^2 \int
{d^3k \over (2\pi)^3} {\sin^2 {\tilde{p}\cdot k\over  2} \over k} 
\left( n_B(k) + n_F(k) \right) + g^2 P^2 \bar{\Sigma} \,.} 
Here $n_B$ and $n_F$ are respectively the Bose-Einstein and Fermi-Dirac 
distributions.
Since the scalar fields are massless at tree-level,  the second term
is $O(g^4)$ and can be dropped in the sequel.
The self-energy \selfscalar\  then leads to the following dispersion relation
\eqn\dispscalar{ p_0^2 = p^2 + 2g^2 T^2 - 
{4 g^2 T\over \pi |\tilde{p}|} \tanh{\pi |\tilde{p}| T\over 2} \, ,}
A plot of \dispscalar\ for different temperatures is shown in
figure 1. 
\ifig\bumps{
Dispersion relation for scalars in \nfour\ Yang-Mills for different temperatures.
The momentum is taken to lie entirely in the non-commutative directions.
The dashed line shows the light-cone. The dotted line
shows the momentum below which the group velocity 
is bigger than one.}{
\epsfxsize=3.8truein\epsfysize=2.3truein
\epsfbox{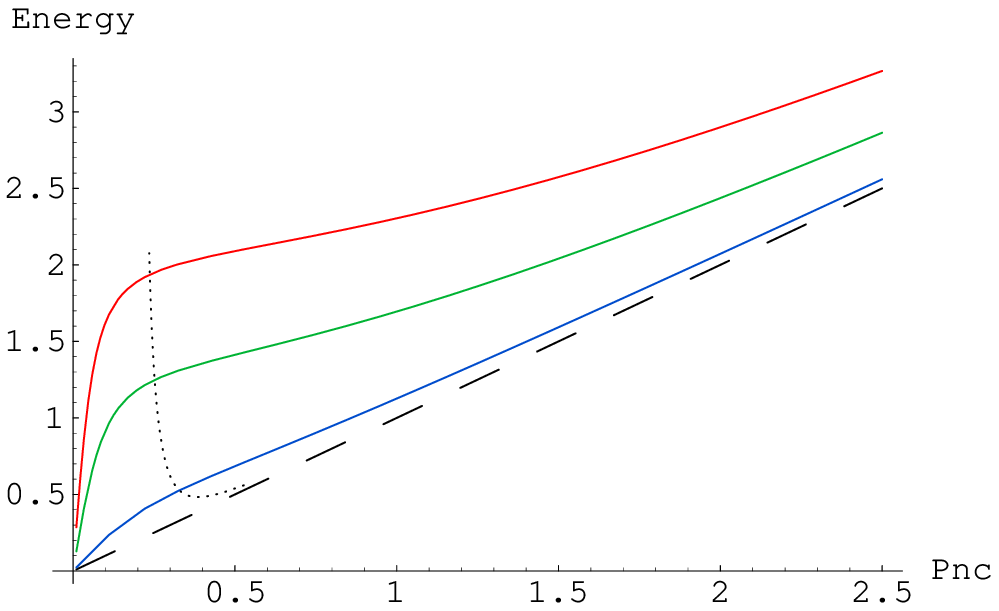} 
}

For large $\pnc$ the dispersion relation is that of a massive particle. 
The last term
in \dispscalar\ corresponding to the non-planar contribution to the 
self-energy is suppressed. The planar contribution gives a thermal mass to the scalar modes,
$m^2_T = 2 g^2 T^2$. 
This is like in ordinary field theories at
finite temperature where particles
are dressed by interactions with the thermal bath and (usually) get a
temperature dependent mass.
At small $\pnc$ however, the planar and non-planar contributions tend to cancel
each other. This is a reflection of the vanishing of the couplings
as $\pnc$ goes to zero. It might be useful to think of the
non-commutative particles as dipoles or rigid rods in the non-commutative 
plane, of size $l \sim \theta \pnc$,
orthogonal to $\pnc$ and
with only the
endpoints interacting \bigsus. As $\pnc \rightarrow 0$, the dipoles are non-interacting.
It immediately follows from these  considerations 
that there must be a domain of $\pnc$ with group velocity 
$\partial p_0/ \partial p  > 1$. We expect this result to be true to all
orders in the weak coupling expansion.

In figure 1, the region corresponding to $\partial p_0/ \partial p  > 1$ is to the left
of the dotted line. For small momenta lying entirely
on the non-commutative plane, $p_0 = c_0 p - 
\gamma p^3 + O(p^5)$ with $c_0 = \sqrt{1+{g^2 \pi^2 \theta^2 T^4 \over 6}}$
and $\gamma = {g^2 \pi^4 \theta^4 T^6\over 120 c_0}$. This is the
dispersion relation of the linearized Korteweg-de Vries equation and it can be shown that the profile of a scalar disturbance travelling along the
non-commutative directions is given in terms of an Airy function:
${1\over 2 (2\gamma t)^{1\over 2}} Ai\left( {x-c_0 t\over (3\gamma t)^{1\over3}}\right)$. As discussed in \we\
the first crest of the wave-train is well defined outside the light-cone for 
large times $t>\sqrt{\gamma\over (c_0-1)^3}$. 

In \we, we argued that
superluminous wave propagation does not {\it a priori} violate 
causality because of the
existence of a preferred class of reference frames where time can be
considered as a commutative coordinate. 
Of course, the reason why group velocities larger than one are possible is
that the theory is not Lorentz invariant. However
this is not seen at tree level in perturbation theory since the
free part of a non-commutative field theory does not
differ from its commutative counterpart. 
The violation of Lorentz symmetry might be more directly noticeable in other
sectors of non-commutative field theories. 
A good example is provided by non-commutative solitons, classical
solutions of the equations of motion which exist only because the 
non-commutativity scale prevents them from collapsing \strominger. It has been shown
recently that these solitons can travel at  
velocities larger than the speed of light \aki\park.

There is another facet to the issue of superluminosity. \nfour\ non-commutative Yang-Mills
is the effective field 
theory  on a $D3$-brane in presence of a large NS $B$ background field \ncall. 
In this framework, and as explained in \sw, one must distinguish between the 
open string $G_{\mu\nu}$ and closed string $g_{\mu\nu}$ metrics.
 In the Seiberg-Witten limit, which can be defined as $\alpha^\prime B \rightarrow \infty$ for fixed $g_{ij}$ with $i,j$ spatial, 
$$
G_{ij} \sim (\alpha^\prime B)^2 g_{ij} \, .
$$
Then $g_{ij}/G_{ij} \rightarrow 0$ and the speed of light in the open string metric effectively
 goes to zero. Conversely, for fixed $G_{ij}$ and $g_{ij}/G_{ij} \rightarrow 0$, the speed of light in the closed
 string metric effectively goes to infinity. In either case, events which
 from  the open string metric point of view propagate faster than the speed of light are
 always in the future light-cone 
of the closed string metric and there is no {\it per se} superluminal propagation\foot{We thank D. Bak for discussions of this issue. (See also \aki.)}. 
Nevertheless, for fields on the D-brane, there is no limit to the speed of light in the
non-commutative plane. 
Such a phenomenon is not specific to the non-commutative gauge theory/string
 setting but has been discussed for instance in the context
of asymmetrically warped Randall-Sundrum
cosmological scenarios  \csaba\langlois\foot{There has also been some work 
on a brane world scenario with  non-commutative extra dimensions (but without 
gravity) \riotto. 
In this setting we live on a soliton of non-commutative space, like in \strominger. 
Space-time on the soliton is Poincar\'e invariant, but
the bulk is non-commutative and    
signals could in principle travel at arbitrary large velocities. 
}.

In the following sections we will extend our discussion to the fermion and  gauge boson sectors
of 
\nfour\ non-commutative $U(1)$ Yang-Mills theory. Not surprisingly, we will 
find that 
interactions give rise again to group velocities larger than one.
But we will also uncover something new. We will 
show that, above some critical temperature $T_c$, \nfour\ non-commutative $U(1)$ Yang-Mills theory
becomes thermodynamically unstable because of the appearance of a tachyonic 
collective mode.

\newsec{Vector Quasi-Particles and Tachyons}

We want to calculate the 
photon polarization tensor at finite temperature and at one-loop.
This issue is technically involved already for ordinary gauge theories.
Physically, this is because the polarization tensor at finite temperature 
encompasses many different
phenomena, like Debye screening of static electric charges, generation
of a thermal mass for the transverse modes, appearance of collective
longitudinal excitations and the physics of Landau damping \kapusta\bellac.
These are  soft phenomena, relevant on scales larger than  the typical particle wavelength in the thermal bath.  
For soft external momenta, $p,\omega \ll T$,
the leading contribution to $\Pi_{\mu \nu}(\omega,p)$ 
is due to hard internal 
momenta, $k \sim T$, and the calculations can be greatly simplified by expanding the integrand 
in powers of $p/k$. Keeping only the leading
term is known in the literature as the hard thermal loop (HTL) approximation \rob.  
In the non-commutative case, the non-planar contribution to the polarization 
tensor involves three independent parameters: $p$, $\tilde p$ and $T$.
Two of them,  $1/\tilde p$ and the temperature $T$, act as
competing cut-off scales. Thus the leading contribution to the non-planar 
diagrams comes from modes $k \sim min(T,1/{\tilde p})$. 
If we expand the integrand in powers of $p/k$ (while keeping the
full dependence on $\tilde p$), the condition for
the first term to dominate the expansion is 
$p\ll min(T,1/\sqrt\theta)$.
The non-commutative version of the HTL approximation applies thus for 
external momenta smaller than both the temperature and the
inverse Moyal cell radius. Our aim is to study effects associated
with the UV/IR mixing characteristic of non-commutative theories.
Since the HTL approximation takes into account the hard momenta
circulating in loops, it will also encode the main contribution 
to UV/IR mixing.
A non-trivial point is that in the HTL approximation the correction to the
polarization tensor (in fact to all two-points functions) 
is gauge independent \rob. This is a well-known result in ordinary 
gauge theories at finite temperature \rebhan\ and its generalization to non-commutative theories
is straightforward. We give an example of this in the appendix B. Let us remind that 
in ordinary gauge theories at zero temperature, 
the poles of propagators are gauge invariant quantities \hooft. In non-commutative gauge theories, 
there is no general proof of this statement, but only checks at one-loop order \ruiz\petriello\foot{At finite temperature, in the HTL loop approximation, the result is a bit stronger, as the whole correction to
the propagator  
is gauge independent, thus even off mass-shell.}.
This is nevertheless sufficient to extract physical information from otherwise 
gauge dependent two-point functions.  
In particular, the poles of 
the resummed propagators (and their residues at
the poles) are gauge invariant quantities and should correspond to physical excitations of the system
 at weak coupling.

The photon polarization tensor in the HTL approximation is given by 
\eqn\pola{
\Pi_{\mu \nu}(\omega, p) = 32 \, g^2  \int {d^3 k \over (2 \pi)^3} 
 \left[ {d n(k) \over dk} \left( 
{i \omega \over P . {\hat K}} - 1 \right) {\hat K}_{\mu} {\hat K}_{\nu}
+ {n(k) \over k} \; ({\hat K}_{\mu} {\hat K}_{\nu}+a_{\mu \nu}) \right]
\sin^2 {{\tilde p} . k \over 2} \, ,}
where ${\hat K}\!=\!(-i,{\bf {\hat k}})$ and $P\!=\!(-\omega,{\bf p})$, 
$n\!=\!n_B+n_F$ and $a_{\mu\nu}\!=\!diag(1,-1,-1,-1)$.
Note that the analytic continuation of the discrete Matsubara frequencies 
to real continuous external frequencies $\omega \rightarrow - i p_0$ is trivial in \pola.
In the limit of soft external momenta, \pola\ can be evaluated by expanding 
$\sin^2 {{\tilde p}.k \over 2}$. The resulting 
expressions are presented in Appendix A. In this section, we 
concentrate on the first non-trivial contribution at small momentum, 
which is ${\cal O}({\tilde p}^2)$ \foot{More precisely, the leading 
term at small momentum is ${\cal O}(g^2 \theta^2 T^4 p^2)$. 
Corrections come in two forms. There are higher terms in the low 
momentum expansion,  
${\cal O}(g^2 T^2 (\theta T p)^{2k})$, which are negligible 
when $\theta T p\!<\!<\!1$. There can be also higher loop corrections 
to the leading term, 
but power counting shows that they are ${\cal O}(g^{2 k} \theta^2 T^4 p^2)$,
and thus small at weak coupling.}. We have  seen in 
the previous section
that the most interesting modifications of the spectrum are likely 
to occur for very low momenta. Also, from now on, we take  $p$ to lie  
entirely in the non-commutative plane.

\ifig\plasmons{The figure shows the slope $x={p_0\over \pnc}$ at 
$\pnc=0$ of the 
dispersion relation for
transversal and longitudinal vector excitations as a function of $T$. 
The temperature is given in units of $1/\sqrt{g \theta}$.
Longitudinal vector excitations exist only for
$T\geq T_c$.  
}
{
\epsfxsize=3.8truein\epsfysize=2.3truein
\epsfbox{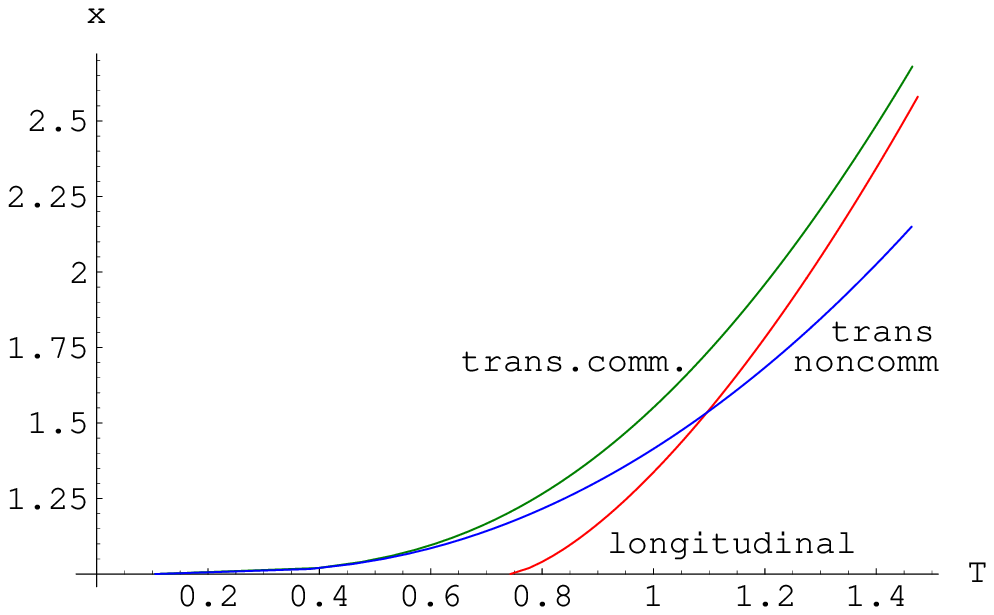}
}

In an ordinary plasma, the longitudinal component of the gauge field can propagate due to the existence of
collective charge oscillations, so-called plasmons. 
The longitudinal components of the polarization tensor can be written
as $\Pi_{\mu \nu}^L=F P_{\mu \nu}^L$ where $P_{\mu \nu}^L$ is 
the longitudinal projector $(P^L)^2 = P^L$ defined by 
$$
P_{\mu\nu}^L = -(g_{\mu\nu} - P_\mu P_\nu/P^2) - P_{\mu\nu}^T
$$
and $P^T$ is the transversal projector
$$
P_{00}^T = P_{0i}^T =0\;\;\; P_{ij}^T = \delta_{ij} - \hat p_i \hat p_j
$$ 
The poles of the propagator coming from the longitudinal sector 
are then given by 
\eqn\dphL{
P^2-F= P^2 \left[\, 1 \, +\, {\pi^2 g^2 \theta^2 T^4 \over 3} \left(
2  - 3 x^2 +{3 \over 2} \, x \, (x^2-1) \log {x+1 \over x-1}
\right) \right]=0 \, .}
where $x = i \omega/p = p_0/p$.  
The solution $P^2\!=\!0$ is a gauge artifact due to the covariant gauge used in the calculation.
The zeroes of the expression in square brackets represent
the true physical excitations of the system. 
The function multiplying $\pi^2 g^2 \theta^2 T^4/3$ inside the square 
brackets is even in $x$, real for $x^2>1$ and
monotonically increasing from $-1$ to zero for $x>1$. 
Therefore there are solutions to \dphL\ only provided   
$T\geq T_c$, where the critical temperature is given by
\eqn\critT{
T_c=\left( {3\over \pi^2 g^2 
\theta^2}\right)^{1\over 4} \, .} 

In a usual plasma of massless particles, the plasmon mode exists at any non-zero temperature.
 An unusual 
feature of the non-commutative plasma is that there is a critical temperature, 
below which the longitudinal excitations do not exist. 
We can shed some light on this phenomenon by thinking
of the system as a gas of dipoles of size $l \sim \theta T$.
On scales $p >1/\theta T$, the system  
looks like a gas of charged particles. Now, ordinary 
plasmons exists only  for soft momenta $p \lsim g T$ \bellac. 
When $1/\theta T< gT$
there is then a range of momenta large enough for resolving the dipoles 
into individual charged particles
but still lower than the  characteristic scale of gauge interactions, $gT$.
In this situation, one would expect propagating longitudinal modes. 
On the other hand, when $1/\theta T> gT$ 
the system essentially behaves as a gas of neutral particles
and there is no reason for the plasmons to exist. From these considerations
we expect the appearance of plasmons at a temperature $T\sim 
{1/ \sqrt{g \theta}}$, which is indeed what we found from \dphL.
We should emphasise that \dphL\ is only valid for $p<1/T\theta$
and the only information we can reliably derive is the group velocity at 
$p=0$. In particular, for $T=T_c$, $x=1$, and then it increases 
monotonically with 
the temperature. From this we conclude that non-commutative plasmons are 
massless and superluminous at the low end of the spectrum. For  
$T \gg T_c$, the dispersion relation of the plasmons is
\eqn\plas{
p_0 \approx \sqrt{2 \over 15} \pi g \theta T^2 \,p .}

At higher momenta, we expect that the plasmon spectrum behaves 
qualitatively as in \bumps. This is because for large momenta, 
$\tilde p\cdot k \sim p \theta T \gg 1$, the 
non-planar contribution to the polarization tensor is suppressed 
compared to the planar one. The polarization tensor reduces to 
$$
\Pi_{\mu \nu}(\omega, p) \approx 4 g^2 T^2 \int {d\Omega \over 4 \pi}  
\left({i \omega \over P . {\hat K}} {\hat K}_{\mu} {\hat K}_{\nu} + \delta_{\mu 4}\delta_{\nu 4}\right )
$$
For $p \ll g T$ but $p \theta T \gg 1$,  the
plasmon dispersion relation becomes
$$
p_0^2 = {3 \over 5} p^2 + {4\over 3} g^2 T^2 ,
$$
like in an ordinary plasma.
Finally, we should notice that the Debye mass vanishes in the 
non-commutative plasma, $\Pi_{\mu\nu}(0,p)
\rightarrow 0$ as $p\rightarrow 0$ so that a static background electric field 
is not screened \miguel.

\bigskip 

We now turn to the dispersion relations of the transverse photons in the 
thermal bath. Since we take the momentum to lie entirely in the
non-commutative plane, one of the transversal directions
is orthogonal to it, and thus local, while the 
other lies on the non-commutative plane, along $\tilde p$.
Concentrating again on the small momentum limit, 
the dispersion relation can be written as
\eqn\dphT{
x^2-1 =   {\pi^2 g^2 \theta^2 T^4 \over 3} \left[\, {c_i \over 4} \, 
\left( 5 x^2 -3 x^4 + {3 \over 2} x (x^2-1)^2  \log {x+1 \over x-1} \right)
\,-\, \delta_{2i}\, \right]  \, ,}
where $i=1,2$ refers to commutative and non-commutative transverse 
directions respectively. The constants
in \dphT\ are $c_1\!=\!1$, $c_2\!=\!3$. Since non-commutativity breaks 
rotational invariance, the dispersion relation is different for each 
polarization. The {\it rhs} is again an even 
function of $x$ and is real for $x^2>1$. It is positive and
monotonically decreasing for $x \in (1,\infty)$, implying that there is 
always a unique solution to \dphT\ in that interval. For low
temperatures, the dispersion relation can be approximated by
\eqn\dstemp{
p_0 \approx \left (1+{\pi^2 g^2 \theta^2 T^4 \over 12}\right) \, p ,
}
for both transverse photons (it is interesting to note that 
\dstemp\ coincides with the result for the
scalars modes).
For high temperatures, the group velocity is different for the two 
polarizations,
\eqn\dhtemp{
p_0 \approx \sqrt{5-c_i \over 30} \pi g \theta T^2 \,p .}

These modes simply correspond to the $T=0$
transverse photon, dressed by the interactions with the thermal bath. 
Interestingly, they are not the only  solutions of the dispersion 
relation \dphT. Replacing $x \rightarrow i x$ in \dphT\ we get
\eqn\tachy{
x^2+1 = {\pi^2 g^2 \theta^2 T^4 \over 3} \left[ {c_i \over 4} 
\left( 5 x^2  +3 x^4 - 3 x (x^2+1)^2 \arctan {1 \over x} \right)
\, + \, \delta_{i2} \, \right] \, .}  
The first term of the {\it rhs} is always negative for $x$ real.
Therefore there is no solution for the photon polarized perpendicular 
to the non-commutative plane.
For the photon polarized along the non-commutative directions however,  
the term in square brackets equals one at $x=0$
and decreases monotonically with $x$. This implies that for 
$T>T_c$, with $T_c$ given by \critT, there is a purely imaginary solution 
to the dispersion relation \dphT. Thus we find a new transverse 
collective excitation
with imaginary energy and, hence, tachyonic in nature.
It is interesting that the critical temperature $T_c$ above which 
the tachyonic solution
exists equals the one for the existence of the plasmon mode.
Close but strictly above $T_c$, 
the tachyon dispersion relation is
\eqn\tachyon{
p_0 \approx \pm i\; {8 \over 3 \pi} \; \left ({T/T_c}-1\right )\; \vert p\vert
}

Let us have a closer look at the polarization tensor \pola.
It can be divided in two pieces $\Pi_{\mu \nu}\!=
\!\Pi_{\mu \nu}^1\!+\!\Pi_{\mu \nu}^2$. The first one is given by 
\eqn\pone{
\Pi_{\mu \nu}^1= 32 \, g^2  \int {d^3 k \over (2 \pi)^3} 
{d n \over dk} \left( 
{i \omega \over P . {\hat K}}{\hat K}_{\mu} {\hat K}_{\nu}  + \delta_{4 \mu}
\delta_{4 \nu} \right) \sin^2 {{\tilde p} . k \over 2} \, .}
This expression is analogous to the one of an ordinary plasma, 
except for the $\sin {{\tilde p}\cdot  k \over 2}$, which acts as
momentum dependent form factor. This piece is responsible
for the first term in the {\it rhs} of \dphT. The second contribution to
the polarization tensor is
\eqn\ptwo{
\Pi_{i j}^2= 32 \, g^2  \int {d^3 k \over (2 \pi)^3} 
\left[ \left( {n \over k} -{d n \over dk}\right) {\hat k}_i {\hat k}_j
-{n \over k} \delta_{ij} \right] \sin^2 {{\tilde p} . k \over 2} \, ,} 
with $\Pi_{44}^2\!=\!\Pi_{4 i}^2\!=\!0$. This piece 
vanishes identically in ordinary QED or Yang-Mills theory at finite T and therefore
genuinely reflects the non-commutativity of the coordinates. 
It can
be evaluated in closed form to give
\eqn\pole{
\Pi_{\mu \nu}^2= 4 g^2 T^2 \left( {\tanh {\pi {\tilde p} T \over 2}
\over \pi {\tilde p} T}-{1 \over 2 \cosh^2 {\pi {\tilde p} T \over 2}}
\right) {{\tilde p}_{\mu}{\tilde p}_{\nu} \over |{\tilde p}|^2} \, .}
Notice that $\Pi_{\mu\nu}^2$ is transverse, because $\tilde p_\mu p^\mu=0$. It is directly analogous
to the new contribution to the polarization tensor specific to non-commutative gauge theories and first derived in \haya\sus.
This term is also at the origin of the delta term in the {\it rhs}
of \dphT, which is essential for the existence of the tachyon collective mode.
 We can not derive a closed form 
expression for the polarization tensor for arbitrary $p$. However
using  $\Pi_{ii}^1(p_0=0,p)\!=\!0$ we observe that
the dispersion relation for $p_0\!=\!0$ reduces to 
$p^2 \delta_{ij}=\Pi^2_{ij}$. This is non-trivial only for the 
polarization along $\tilde p$. For $T > T_c$
that equation has a solution at $p>0$\foot{
The slope of $\Pi^2_{22}$ at the origin
is $T^4/T_c^4$. When $T>T_c$, $p^2-\Pi^2_{22}<0$ at 
small $p$ while it tends to $+\infty$ at high $p$. It is easy to 
show that that function has a single minimum and therefore there is a unique 
point $p>0$, for $T>T_c$, such that $p^2=\Pi^2_{22}$.}. It is natural to
assume that this additional pole belongs to the tachyonic branch.
This suggests the following schematic structure of the
tachyon branch: the branch starts at $p_0\!=\!0$ for $p\!=\!0$; the
square of the energy becomes negative as $p$ increases
until it reaches a minimum; it then increases until it reaches
$p_0\!=\!0$ again, at a point which we will denote $\bar p$. 
For very high temperatures, we have
\eqn\endt{
{\bar p}\approx \left( {4 g^2 T \over  \pi \theta} \right)^{1 \over 3} \, .} 
\ifig\tachyonf{Dispersion relation of the tachyonic branch for 
different temperatures. On the vertical axes we have plotted $p_0^2$.  
}
{
\epsfxsize=3.8truein\epsfysize=2.3truein
\epsfbox{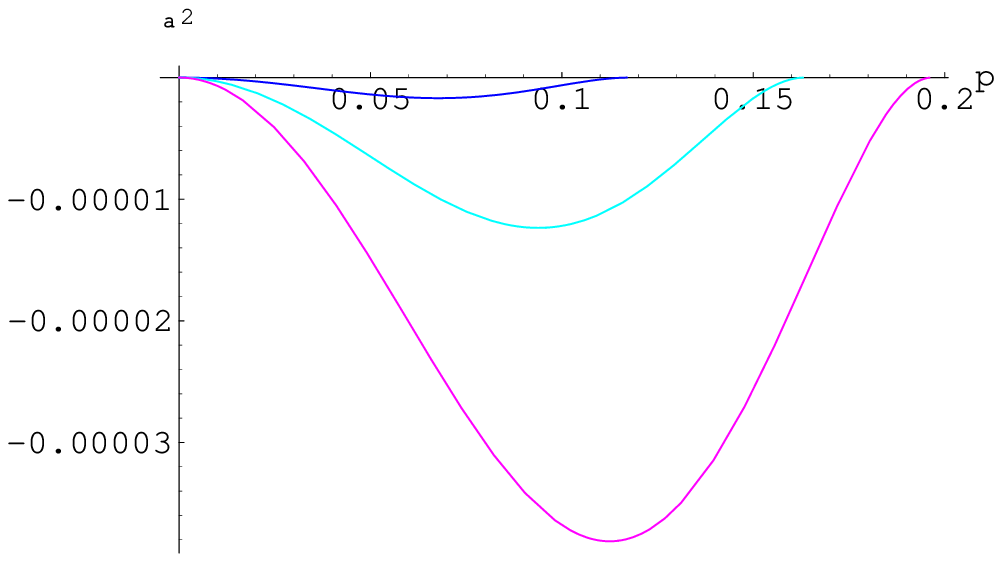}
}

Let us investigate the behaviour of the tachyonic branch of
the dispersion relation around $\bar p$ more closely.
In the neighbourhood $p_0=0$ and when $p_0/p\approx 0$ we find
$\Pi^1_{22}\approx i p_0 sgn(\Im p_0) 
f(p,\tilde p)$ with $f>0$. 
At first order in $p_0$, the dispersion relation reduces to
$p^2=\Pi^2_{22}+\Pi^1_{22}$. This implies that $p_0$ must be 
imaginary and therefore
\eqn\dispbar{
p^2=\Pi^2_{22}-|p_0| f(p,{\tilde p}) \, .}
Since the second term in the {\it rhs} is negative
there are solutions only for $p^2<\Pi^2_{22}$.
Using the arguments of footnote 5, we observe that  
there are no solutions to \dispbar\ for
$p$ bigger that $\bar p$! The polarization tensor has a branch
cut in the complex $x$ plane between $(-1,1)$. This is true 
to all orders in the $\tilde p$ expansion ({\it cf.} the expressions
in appendix A). Thus what happens is that the pole corresponding
to the tachyonic collective mode becomes part of this branch cut
for higher momenta. We have confirmed this behaviour numerically
for different temperatures above the critical one using 
the results of appendix A (see figure \tachyonf).

In retrospect, 
the existence of   tachyonic collective excitations could 
have been expected from the behaviour of non-commutative gauge theories 
without supersymmetry. In these cases, non-planar diagrams give rise 
to pole-like infrared singularities in  gauge boson propagators \haya\sus. 
The sign of these poles depends on the
relative number of fermionic and bosonic degrees of freedom
in the adjoint representation. For the transverse photon polarized 
along $\tilde p$, the dispersion relation in four dimensions is 
\eqn\susw{
p_0^2=p^2 + {c \over \pi^2} {g^2 \over |{\tilde p}|^2} \, ,
}
where $c= 2 N_f -2 - N_s$, with $N_f$ and $N_s$ respectively 
the number of Majorana fermions and real scalars in the adjoint 
representation. 
Thus, if there are more bosons than fermions the infrared singularity tend 
to destabilize the system: $p_0^2 \rightarrow - \infty$ as $p \rightarrow 0$.
Contrary if there more adjoint fermions than bosons, infrared divergences
remove the lower end of the spectrum: $p_0 \rightarrow + \infty$ as 
$p\rightarrow 0$. In supersymmetric theories these infrared singularities
are absent, since $c=0$. However, because the presence of a thermal bath 
breaks supersymmetry, 
it is natural that a behaviour reminiscent of non-supersymmetric 
theories dominates at high  temperatures. In particular
as $T$ increases, and in the static limit, we should recover the physics of a 
three-dimensional Euclidean non-commutative gauge theory. 
In that limit the fermions decouple and only the zero modes
of the bosonic degrees of freedom survive. Using that $\Pi^1_{ij}(p_0=0)
\!=\!0$, it is easy to derive the dispersion relation for the 
three-dimensional photon polarized along $\tilde p$ from our previous 
results. We obtain
\eqn\dphthree{
p^2 \, - \, {4 g_3^2 \over \pi} \, {1\over 
|{\tilde p}|}\, = \, 0 \, ,}
where the three-dimensional coupling constant is given by $g_3^2=g^2 T$. 
We have checked that this coincides with the infrared leading term  
of the dispersion relation at one-loop for a transverse photon
polarized along $\tilde p$ 
in a three-dimensional non-commutative gauge theory with seven 
adjoint scalars. That we only reproduce the leading term is a consequence
of the HTL approximation which gives only the leading terms
in a high temperature expansion. This is analogous to keeping only the
UV leading terms in the Feynman integrals at zero temperature ({\it cf.} \sus).
The seven adjoint scalars correspond to the zero modes 
of the six \nfour\ scalars plus the longitudinal mode of the gauge
field. Note that at one-loop the scalar modes acquire a mass of order $g_3^2 T$
in the $T\rightarrow \infty$ limit and non-zero momentum. In ordinary 
thermal field theories they 
could be integrated out to yield an effective theory consisting only of
gauge bosons. In a non-commutative theory however the scalars contribute
to the UV/IR mixing at the one-loop level. Thus they can not be neglected. 
Indeed the coefficient of 
\dphthree\ shows that the 
scalars do not decouple. 
The fermions on the other hand always decouple in the static limit,
 because their Matsubara
 mass is independent of interactions. 
\ifig\missing{Sumrule for the transversal mode
polarized in the non-commutative direction. The temperature is given in units
of $1/\sqrt{g\theta}$. The contribution from the tachyonic pole is not 
included. 
}
{
\epsfxsize=3.8truein\epsfysize=2.3truein
\epsfbox{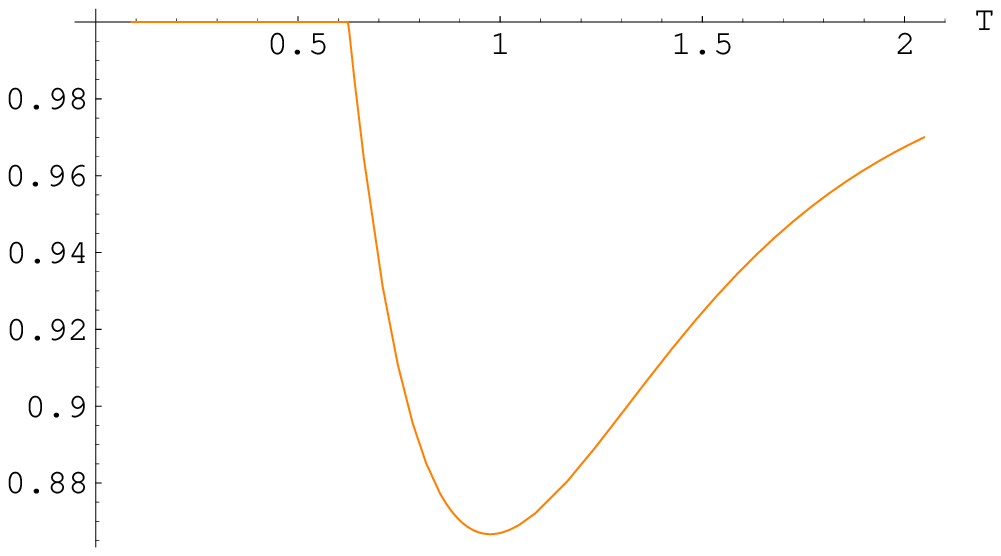}
}
The presence of the tachyon mode can be alternatively detected by 
studying the sum rules of the theory. This will also show us that there is no other
 pole
with $\Im p_0 \neq 0$ than the one we have found. 
We  concentrate on the first non-trivial sum rules, which can be 
derived using standard techniques (see for instance \bellac).
For $p \ll 1/\theta T$ we have
\eqn\srules{\eqalign{
\int_{-1}^{1} {d\xi \over 2 \pi}\, \xi \,\rho_L(\xi)\, + \sum x_k Z_k & =
\, {2 \over 15} \pi^2 g^2 \theta^2 T^4 \, ,  \cr 
\int_{-1}^{1} {d\xi \over 2 \pi} \, \xi \, \rho_T^i(\xi) + \sum x_k Z^i_k\, & =
\, 1 \, ,}}
where the spectral function is $\rho_{L,T}\!=\!-2 \Im (\Delta^{-1}_{L,T})$, 
with $\Delta_{L,T}$  
denoting the dispersion relations
defined in \dphL, \dphT. The residues at the poles are given by
$Z_k = \Bigl(\Bigl.{\partial \Delta_{L,T}\over \partial x}\Bigr)^{-1}
\Bigr|_{x=x_k}$ and $x_k$ denotes the position of the poles.
The integrations in $\xi$ can be thought of as integrations in $p_0$ by
 keeping $p$ fixed.
We have checked the sum rules for the longitudinal modes and the 
transverse modes polarized perpendicular to the non-commutative plane
numerically.
For the transverse modes polarized in the non-commutative plane, 
the sum rule ignoring the tachyonic poles is plotted in
figure \missing.
It is saturated below the critical temperature and shows the missing 
contribution from the tachyonic poles above it. Taking into account
the tachyonic poles also this sumrule is saturated. We leave for 
section 5 a discussion of the physical implications of the tachyonic 
excitations.

We would like to end this section by commenting on some generalizations.
First let us consider the theory with $\;U(N)$ gauge group.  
Due to the Moyal bracket interactions the $\;U(1)$ component
does not decouple from the $\;SU(N)$ modes. 
The polarization tensor at one-loop in the $\;U(1)$ sector 
is now \arm\kh\
$$
\Pi_{\mu \nu,N}= N \, \Pi_{\mu \nu ,1} \, .
$$ 
Therefore all the results in this section apply by just substituting 
$g^2 \rightarrow g^2 N$. On the other hand, the polarization tensor 
for the $\;SU(N)$ modes does not receive non-planar contributions at 
one-loop \mvrs\arm\ and, at that order, their spectrum 
is that of an ordinary theory.

Another generalization is to consider theories with less
supersymmetry. As long as only matter in the adjoint representation
is present we expect qualitatively the same behaviour as we have
found in this section. If we include also matter in the fundamental 
representation some differences will however appear. Matter in the 
fundamental representation only gives rise to planar contributions to 
the polarization tensor at one-loop \haya. Thus, as in ordinary theories, 
there will be a non-vanishing Debye mass and longitudinal
photon excitations for all $T>0$. 
In addition, we expect the tachyonic pole in the non-commutative transverse 
polarization still to be present. This is because in the static limit
and as $T\rightarrow \infty$ one obtains a non-commutative 
non-supersymmetric three-dimensional theory, with tachyonic 
behaviour similar to \dphthree.
In fact there is no new contribution from fundamental matter to
the genuinely non-commutative part of the polarization tensor \ptwo.
Of course the overall coefficient of this term will differ depending on 
the content of adjoint matter, and this will change the
value of the critical temperature above which the tachyon appears.

\newsec{Fermion dispersion relations and plasmino modes}

For completeness, we investigate the finite temperature self-energy of
the fermions. Let us start by briefly recapitulating 
what happens in ordinary field theory. The most striking feature
of the fermion dispersion relations at finite temperature is the appearance
of a new branch of excitations. In addition to the particle-like states already
present at zero temperature,  there are hole-like excitations \plasmi. 
The physical
picture behind this phenomenon is that the anti-fermion annihilation operator 
acts
non-trivially on the equilibrium thermal state by creating  a hole state. 
In the free case, the system with a hole state of momentum $p$ would have 
lower energy by $p$ 
than the original equilibrium thermal state and thus it is not 
accessible. In the interacting case however, it costs an energy 
which typically is ${\cal O}(g T)$
to create a hole and this state can have positive energy provided 
$p \lsim g T$. Therefore hole states might be excited.  

In the non-commutative \nfour\ gauge theory the self-energy of the fermions
in the HTL approximation takes the form
\eqn\selfenergy{ \Sigma = {8 g^2\over \pi^2} \gamma_\mu \int k^3
 n(k) dk \int { d\Omega \over 4\pi} 
{\hat{K}^\mu \over P.\hat{K}} \sin^2{\tilde{p}.\vec{k}\over 2}\,.}
\ifig\plasminos{The slope $x={p_0\over\pnc}$ at $\pnc =0$ for the
fermionic excitations. The temperature is given in units of 
$1/\sqrt{g\theta}$. The plasmino mode appears only above the critical 
temperature $T_c^f$.
}
{
\epsfxsize=3.8truein\epsfysize=2.3truein
\epsfbox{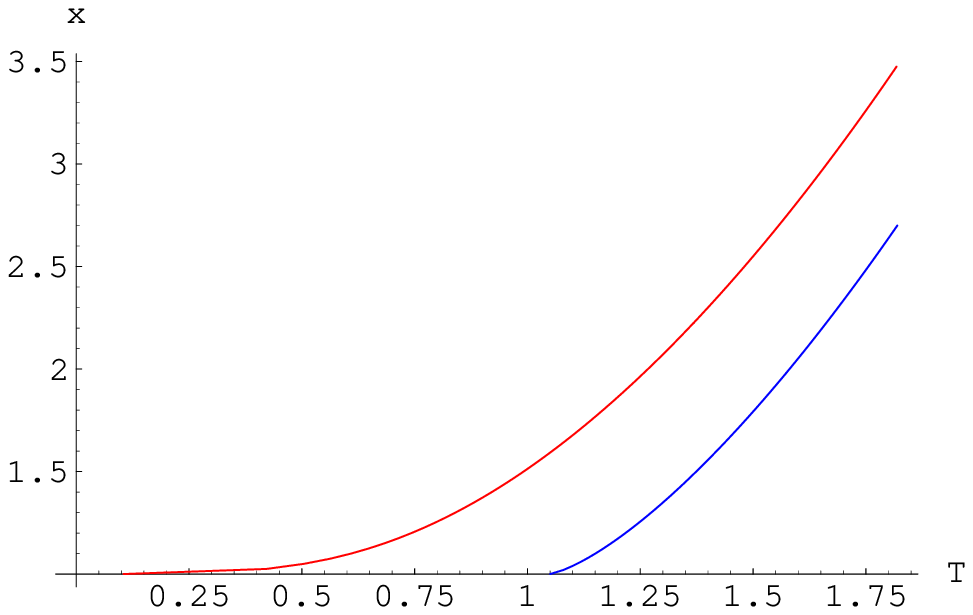}
}
Again we  concentrate on the lowest order contribution in $\tilde p$ from
the series expansion of the sine. The inverse propagator can be written in the form
\eqn\invpropfermions{\gamma_\mu P^\mu - \Sigma= p \left(\Delta_+ (\gamma_0 - \vec{\gamma}.\hat{p}) +
\Delta_- (\gamma_0 + \vec{\gamma}.\hat{p})\right)\,,}
where
\eqn\dispfermions{\Delta_\pm = x \mp 1  +  {\pi^2 g^2 \theta^2 T^4 \over 8}
\left( x \pm {1\over 3} \mp (1-x^2)(1-{(x\mp 1)\over 2} \log{x+1\over x-1}
)\right)\,.}

Notice that $\Delta_+(-x)= -\Delta_-(x)$. Since we are dealing with 
Majorana
fermions, negative energy solutions can be identified with positive energy
solutions. Thus we look for real positive zeroes of $\Delta_\pm$ as
function of the temperature $T$. In this
region, it turns out that $\Delta_-=0$ has always a solution. For low 
temperatures, it reduces to the vacuum particle-like excitation with 
$p_0 = p $. On the other hand,
$\Delta_+(x)$ has zeroes only for temperatures $T>T_c^f$ with $T_c^f= \left(
{12\over g^2 \pi^2 \theta^2}\right)^{1\over 4}$. 
The situation is thus  similar to the case of the plasmon mode, 
and the plasmino exists only at sufficiently high temperature (see 
\plasminos).

Let us define the residues $Z^i_\pm = \Bigl.\Bigl( {\partial 
\Delta_\pm^{-1} \over
\partial x}\Bigr)^{-1}\Bigr|_{x=x^i_\pm}$, where $x^i_\pm$ denote the zeroes of $\Delta_\pm$. We furthermore define the spectral functions
$ \rho_\pm = -2 \Im ( \Delta_\pm^{-1})$. 
One can easily prove now that the following sum rule holds
\eqn\sumrulefermions{ Z_\pm^1 + Z_\pm^2 + \int_{-1}^1 {d\xi\over 2 \pi} \rho_\pm(\xi) = 1 \,.}
As in the case of the vector bosons, it is implicitly  understood 
that we keep $p$ fixed and small and vary $p_0$ as we vary $\xi$.
Let $x_\pm^1$ be the positive real zero of $\Delta_\pm$. For 
temperatures $T< T_c^f$ there exists only a positive
solution for $\Delta_-$. Thus, for this low temperature region, $Z^1_+$ and
$Z^2_-$ vanish in the sum rule \sumrulefermions. In the low temperature
regime we find $1> Z_-^1 (Z_+^2) > 0.7263$. At very high temperatures
we find $Z_\pm^1 \approx Z_\pm^2 \approx 0.5$. This is reminiscent of
the behaviour of ordinary plasmino modes. Indeed the residues of particle
and hole excitations in ordinary plasmas approach $0.5$ at zero momentum
\plasmi.

Comparing the plasmino and plasmon critical temperatures, 
we observe that $T_c^f>T_c$. Since at $T_c$ a tachyonic 
collective mode also appears, we expect the onset of a phase transition 
at that temperature. In consequence, the relevance of the plasmino mode 
for our system is unclear. 
However the qualitative properties of
the fermion excitations do not depend on the presence of gauge bosons.
We could have studied for example a non-commutative model with
chiral superfields only and Moyal-bracket interactions. 
To give a specific example take 
a model with three chiral superfields and superpotential $W = \Phi^1 * 
\{\Phi^2 , \Phi^3\}_*$. This corresponds to setting the
\none\ vector field in the \nfour\ $\;U(1)$ gauge theory to zero.
The self-energy for the fermions is then $3\over 4$ times the one in 
\selfenergy. The temperature where plasmino modes appear changes
by a factor $\left(3\over 4 \right)^{1/ 4}$.

\newsec{Discussion}

We have investigated the effects of non-commutativity on the
spectrum of gauge theories. One of the most interesting phenomenon 
characteristic of non-commutative field theories is the mixing 
between UV and IR modes. In particular high momentum 
circulating in loops affects physics at arbitrary low energies, and 
generically causes the appearance of IR divergences in the perturbative 
expansion. We have considered a $U(1)$ gauge theory
with \nfour\ supersymmetry. In its ordinary version this theory is
UV finite. Correspondingly its non-commutative deformation is 
free of IR divergences \zan. Finite temperature breaks supersymmetry but 
also provides a natural UV cut-off which allows to study the issue of 
UV/IR mixing in a controlled setting. Accordingly, as the 
temperature is increased, the effects associated 
with non-commutativity become more pronounced.

A first remarkable feature derived from our study of the
dispersion relations is the appearance of superluminous group 
velocities at very soft non-commutative momenta for all degrees of
freedom in the theory. This extends to fermions and vectors the
results derived in \we\ for scalars.
Since Lorentz symmetry is broken there is no group theoretical reason 
for maximum velocity in the non-commutative directions.
As we argued in \we, problems with causality can {\it a priori} be avoided 
in systems with space-like non-commutativity because of the existence of a 
preferred class of reference frames, those where time is a local coordinate
(see also \aki\lp). 

An interesting property of the fermion and vector dispersion relations
is the appearance of collective excitations above a certain critical 
temperature. Hole-like excitations in the fermion sector exists for 
$\pi^2 g^2 \theta^2 T^4\!>\!12$, while collective excitations of the 
gauge bosons appear at a lower temperature $\pi^2 g^2 \theta^2 T^4\!>
\!3$. These are of two types: longitudinal modes and tachyonic modes in 
the transverse sector polarized in the non-commutative plane.
The appearance of propagating longitudinal modes can be understood
by  comparing with the behaviour of ordinary plasmas and picturing 
non-commutative particles as dipoles. In this way 
the very smooth UV/IR effects induced by temperature in the
\nfour\ theory allow to probe the spatial 
extent of non-commutative excitations in a neat way. 

The most striking result of our work is the appearance of tachyonic 
collective excitations above the critical temperature 
$T_c\!=\!(3/\pi^2 g^2 \theta^2)^{1/4}$. This implies that at $T_c$ the 
system becomes thermodynamically unstable and might undergo a phase 
transition. To gain some understanding
on the physics of this instability let us recall that non-commutative 
gauge theories can be alternatively described in terms of ordinary gauge 
fields subject to ordinary gauge transformations \sw. The Seiberg-Witten map,
which relates both descriptions, is given by
\eqn\sw{
F_{\mu \nu}=\left( {1 \over 1- {\hat F} \theta} {\hat F} \right)_{\mu \nu}
+\, \, {\cal O}(\partial {\hat F}) \, ,
}
where $F$ and $\hat F$ denote respectively the ordinary and
non-commutative field strengths. Using \sw\ we can define a new
symplectic form, $\omega=\theta^{-1}+F$, and associate
a new star product to it \wess. Thus, the Seiberg-Witten map allows to 
relate a configuration of the non-commutative gauge field with a 
modification of the Poisson structure that defines the star product.
The tachyonic mode we have found appears precisely in the transverse sector
polarized along a non-commutative direction. Therefore it implies
a growing mode for the field strength ${\hat F}$ with indices on
the non-commutative plane, {\it i.e.} $\hat F_{23} \not= 0$. The previous 
argument then suggests 
that the tachyonic mode corresponds to a destabilization of the 
initial non-commutative structure associated with
$\omega=\theta^{-1}$.

A more physical picture may be obtained from string theory.
The \nfour\ $U(1)$ gauge theory that we have studied can be
realized as the world-volume theory on a D3-brane with an
strong constant magnetic B field or, equivalently, a D3-brane with
an infinite number of delocalized D1-branes bounded to it. 
In particular $B=\theta^{-1}$, which implies an homogeneous 
distribution of D1-brane charge. The appearance of the 
tachyonic mode suggests  that the density of D1-branes becomes inhomogeneous above $T_c$. The new equilibrium
state could correspond to a non-translational invariant distribution
of D1-brane charge, analogous to the stripe phases found in \gubser.
Another possibility is that the phase transition at $T_c$ is
associated to the nucleation of D1-strings, dual to the evaporation of F1-strings  in NCOS \ncos. Although 
attractive, 
the latter possibility seems however unlikely as quite general thermodynamic arguments suggests that this may only 
happen at very large couplings \aki\pr.

Either way, we might expect something non-trivial to happen above $T_c \sim 1/\sqrt{\theta g}$, possibly for 
{\it any} coupling. 
Consider the supergravity dual 
description of \nfour\ non-commutative $U(N)$ theories \akione\maldacena. 
The supergravity background has an interior region that
tends to $AdS_5 \times S^5$, while the region close to the
boundary reproduces the supergravity metric associated
to an infinite set of parallel D1-branes completely delocalized 
in two spatial directions. The value of the radial coordinate $u$
at which the crossover between the two regimes takes place is
$u_c \approx (\lambda /\theta^2)^{1/4}$, with  't Hooft
coupling $\lambda\!=\! 2 g_{YM}^2 N$ and 
 boundary corresponding to $u \rightarrow \infty$.
The field theory at finite temperature corresponds
 to a black hole background in the supergravity dual. 
 It is interesting that the
critical temperature $T_c\approx 1/(g_{YM}^2 N \theta^2)^{1/4}$ 
corresponds to a black hole whose horizon is at $u\approx u_c$.
For $T>T_c$ the supergravity metric does not reduce in
any region to $AdS_5 \times S^5$ and instead looks like
that of smeared D1-branes along all its radial extent.
The temperature $T_c\approx 1/(g_{YM}^2 N \theta^2)^{1/4}$
is the critical temperature for the appearance of tachyon modes at weak 
coupling 
in \nfour\ $U(N)$ non-commutative theory.\foot{Although we do not expect the 
supergravity dual to capture all aspects of a $U(1)$ non-commutative gauge 
theory, it is interesting to note that at large temperatures, $T > T_c$, 
the 
supergravity approximation is valid even for small $N$ provided 
$g_{YM} > 1$ \aki.}
It is thus plausible  that in the $T\!\!-\!\!\lambda$ plane of the phase 
diagram of NCSYM
there is a new line of phase transitions, starting at weak coupling and 
possibly extending in the 
supergravity region. The order and  nature of the phase transitions can not 
be established within our
framework and we leave these questions for future studies.

\vskip1cm

\centerline{\bf Acknowledgments}
\vskip2mm
We would like to thank L. Alvarez-Gaum{\'e}, D. Bak, J. Barb{\'o}n, C.-S. Chu,
R. Emparan, R. Gopakumar, H. Grosse and C. Manuel for discussions.

\appendix{A}{}
Here we collect the results from integrating the series expansion in
$\tilde p$ of the polarization tensor and the fermion self-energy. 
We assume that the momentum lies entirely in the 
non-commutative directions $\vec{p}=\vec{p}_{nc}$. 
The polarization tensor can then be decomposed 
\eqn\decompoltens{ \Pi_{\mu\nu} = F P^L_{\mu\nu} +
G^c P^{T,c}_{\mu\nu} + G^{nc} P^{T,nc}_{\mu\nu}\, .}
Here $P_{\mu\nu}^L$ is the usual
longitudinal projector, $P^{T,c}_{\mu\nu}$ and $P_{\mu\nu}^{T,nc}$ are
the transverse projectors in the commutative and non-commutative directions
respectively. Terms containing ${1\over P.\hat{K}} \sin^2{\tilde{p}k\over2}$
are evaluated by expanding in $|\tilde{p}|$.
Using the integral representation for the hypergeometric function
\eqn\inthyp{ {\Gamma(c)\over \Gamma(a) \Gamma(c-a)}\int_0^1 t^{a-1} (1-t)^{c-a-1} (1- z t)^{-b} = F(a,b,c;z)\,,}
we find for the components of the polarization tensor of the gluons
\eqn\poltensor{\eqalign{
F =&  {8 g^2 T^2\over\pi^2}{P^2|\over p_0^2} \sum_{n=1}^\infty A_n  
F(1,{3\over 2},n+{5\over 2};{\vec{p}^2\over p_0^2})\, ,\cr 
G^{T,c} =& {8 g^2 T^2\over\pi^2} \sum_{n=1}^\infty A_n 
F({1\over 2},1,n+{5\over 2};{\vec{p}^2\over p_0^2})\, ,\cr
G^{T,nc} =& 2 g^2 T^2 \left[ {2 \tanh {|\tilde{p}|  
\pi T \over 2}\over |\tilde{p}| \pi T} - {1\over\cosh^2{|\tilde{p}|  \pi T \over 2}} \right] + 
 {8 g^2 T^2 \over \pi^2}  \sum_{n=1}^\infty A_n (2n+1) 
F({1\over 2}, 1, n+ {5\over 2}; {\vec{p}^2\over p_0^2})\, ,
}}
where we defined 
\eqn\ans{ A_n = (-)^n {2n+2\over 2n+3} \,
{2^{2n+2}-1\over 2^{2n+1}}\, \zeta(2n+2)\,(|\tilde{p}| T)^{2n}  \,.}

The self-energy of the fermions can be obtained in an analogous manner.
\eqn\selfenergy{ \Sigma = {4 T^2 g^2 \over \pi^2} \sum_{n=1}^\infty B_n 
\left[ \gamma_0\; {F({1\over 2},1,n+{3\over 2};{\vec{p}^2\over p_0^2})\over p_0} - 
(\vec{\gamma}.\hat{p})\; {|\vec{p}| F(1,{3\over 2}, 
n+{5\over 2}; {\vec{p}^2\over p_0^2})\over p_0^2 (2n+3)} \right]\, ,}
and
\eqn\bns{ B_n = (-)^n\, {2^{2n+2}-1\over 2^{2n+1}}\, \zeta(2n+2)\,
 (|\tilde{p}| T)^{2n}\,.}

\appendix{B}{}
We want to show that the two-point functions of non-commutative gauge theories are
 gauge invariant in the HTL approximation. The argument is actually a straightforward generalization
of the results of \rob. For convenience  we repeated here for the simplest case of the fermion self-energy. 
We refer to \rob\ or \rebhan\ for a discussion of the gauge independence 
of the polarization tensor. 
 
In a covariant gauge, the gauge-dependent contribution to the self-energy is of the form
$$
\delta_\xi \Sigma(P) \sim \int_K \gamma_\mu {1\over \Pslash{P}-\Pslash{K}}
 \gamma_\nu {\cal D}^\xi_{\mu\nu}(K) \sin^2 {\tilde p\cdot k\over 2}\,.
$$
With ${\cal D}_{\mu\nu}^\xi = K_\mu K_\nu/K^4$ ,
$$
\delta_\xi \Sigma(P) \sim \int_K \Pslash{K}(\Pslash{P} - \Pslash{K})\Pslash{K}\,{1\over (P-K)^2} {1\over K^4}\sin^2 
{\tilde p\cdot k\over 2}\,.
$$
With $\Pslash{K} (\Pslash{P} - \Pslash{K}) \Pslash{K} = - (P-K)^2 \Pslash{K} + P^2 \Pslash{K} - K^2 
\Pslash{P} \approx - (P-K)^2 \Pslash{K} - \Pslash{P} K^2$, using $P \ll K$,  the gauge-dependent part reduces to 
$$
\delta_\xi \Sigma(P) \sim - \int_K \Pslash{P} {1\over (P-K)^2}{1\over K^2}\sin^2 
{\tilde p\cdot k\over 2}\,.
$$
For soft ($P \ll T$)  and small ($p \ll 1/\theta T$) external momentum, this contribution is manifestly  subdominant compared to the HTL contribution to the self-energy,
$$
\Sigma_{HTL}(P) \sim \int_K \Pslash{K} {1\over (P-K)^2} {1\over K^2}\sin^2 
{\tilde p\cdot k\over 2}\,.
$$
For soft ($P \ll  min(T,1/\sqrt{\theta})$) but otherwise generic $\tilde p\cdot k$,
 because $\sin^2 x \leq 1$ the (absolute value of the) integrands in both expression are bounded
 above by the corresponding expressions in the commutative version of the theory.
Furthermore, these expressions are  ``in phase'' for the same $p$ and consequently
 the gauge dependent piece is always subleading compared to the HTL contribution.

\listrefs

\end